\documentclass[12pt]{article}
\begin{document}

\title{The nonlinear directional coupler. An analytic solution}
\author{R.Vilela Mendes\thanks{%
vilela@cii.fc.ul.pt} \\
{\small Universidade T\'{e}cnica de Lisboa e Grupo de F\'{i}sica-Matem\'{a}%
tica, }\\
{\small Complexo Interdisciplinar,} {\small Av. Prof. Gama Pinto 2, 1649-003
Lisboa, Portugal}}
\date{}
\maketitle

\begin{abstract}
Linear and nonlinear directional couplers are currently used in fiber optics
communications. They may also play a role in multiphoton approaches to
quantum information processing if accurate control is obtained over the
phases and polarizations of the signals at the output of the coupler. With
this motivation, the constants of motion of the coupler equation are used to
obtain an explicit analytical solution for the nonlinear coupler.
\end{abstract}

\section{Introduction}

Directional couplers are useful devices currently used in fiber optics
communications. Because of the interaction between the signals in the input
fibers, power fed into one fiber is transferred to the other. The amount of
power transfer can be controlled by the coupling constant, the interaction
length or the phase mismatch between the inputs. If, in addition, the
material in the coupler region has nonlinearity properties, the power
transfer will also depend on the intensities of the signals \cite{Jensen} 
\cite{Silberberg}. A large number of interesting effects take place in
nonlinear directional couplers \cite{Stegeman} \cite{Stegeman2} \cite{Kenis} 
\cite{Liu} with, in particular, the possibility of performing all classical
logic operations by purely optical means \cite{Wang}. They may also play a
role in quantum information processing. 

The use of the intensity-dependent phase shifts associated to the Kerr
nonlinearity was, in the past, proposed for the construction of quantum gates%
\cite{Milburn} \cite{Glancy} \cite{Fu}. However they rely on one-photon
processes and therefore would require very strong nonlinearities, not
available in the low loss optical materials. On the other hand, the quantum
computation scheme based on linear optics of Knill, Laflamme and Milburn is
probabilistic and relies on a delicate sensitivity of one-photon detectors.
For this reason multiphoton approaches have been explored based either on
the quantumlike behavior\cite{Margarita} \cite{QL1} \cite{QL2} of optical
modes on a fiber\cite{Manko} or on coherent states\cite{Ralph}. For light
beams on a fiber, sizable nonlinear effects are easy to achieve with
available materials. In particular the directional coupler might provide an
already available tool for the implementation of linear or nonlinear gates%
\footnote{%
There have been some speculations\cite{Lloyd} that nonlinear quantum(like)
effects might endow quantum computation with yet additional power.  }.

For quantum information purposes one would require accurate information on
the phases and polarizations of the signals at the output of the coupler.
Analytic solutions are ideal for this purpose although, in general,
difficult to obtain for nonlinear systems. Here, by exploring the constants
of motion of the coupler equation, an explicit analytical solution is
obtained for the nonlinear coupler.

\section{An analytic solution}

Consider two linear optical fibers coming together in a coupler of nonlinear
material. The equation for the electric field is 
\begin{equation}
\triangle E-\mu _{0}\varepsilon _{0}\frac{\partial ^{2}E}{\partial t^{2}}%
=\mu _{0}\frac{\partial ^{2}P_{L}}{\partial t^{2}}+\mu _{0}\frac{\partial
^{2}P_{NL}}{\partial t^{2}}  \label{B.1}
\end{equation}
$P_{L}$ and $P_{NL}$ being the linear and nonlinear components of the medium
polarization. 
\begin{equation}
P_{L}\left( r,t\right) =\varepsilon _{0}\chi ^{(1)}E\left( r,t\right) 
\label{B.2}
\end{equation}
For symmetric molecules (like SiO$_{2}$) the leading nonlinear term is 
\begin{equation}
P_{NL}\left( r,t\right) =\varepsilon _{0}\chi ^{(3)}\left| E\left(
r,t\right) \right| ^{2}E\left( r,t\right)   \label{B.3}
\end{equation}
where an instantaneous nonlinear response may be assumed (except for
extremely short pulses) because in current fibers the electronic
contribution to $\chi ^{(3)}$ occurs on a time scale of $1-10$ fs.

Separating fast and slow (time) variations 
\begin{equation}
\begin{array}{lll}
E\left( r,t\right)  & = & \frac{1}{2}\left\{ \mathcal{E}\left( r,t\right)
e^{-i\omega _{0}t}+c.c.\right\}  \\ 
P_{NL}\left( r,t\right)  & = & \frac{1}{2}\left\{ \mathcal{P}_{NL}\left(
r,t\right) e^{-i\omega _{0}t}+c.c.\right\} 
\end{array}
\label{B.4}
\end{equation}
and using Eqs.(\ref{B.3}) and (\ref{B.4}) one obtains for the $e^{-i\omega
_{0}t}$ part of a transversal mode 
\begin{equation}
P_{NL_{1,2}}\left( r,t\right) =\frac{3\varepsilon _{0}}{8}\chi ^{(3)}\left\{
e^{-i\omega _{0}t}\left[ \left( \left| \mathcal{E}_{1,2}\right| ^{2}+\frac{2%
}{3}\left| \mathcal{E}_{2,1}\right| ^{2}\right) \mathcal{E}_{1,2}+\frac{1}{3}%
\mathcal{E}_{2,1}\mathcal{E}_{2,1}\mathcal{E}_{1,2}^{*}\right] +c.c.\right\} 
\label{B.5}
\end{equation}
The labels $1$ and $2$ denote two orthogonal polarizations.

The dependence on transversal coordinates $\left( x,y\right) $ is separated
by considering 
\begin{equation}
E_{k}^{(i)}\left( r,t\right) =\Psi _{k}^{(i)}\left( x,y,z\right) e^{i\beta
_{i}z}e^{-i\omega _{0}t}  \label{B.6}
\end{equation}
$\Psi _{k}^{(i)}\left( x,y,z\right) $ being an eigenmode of the coupler with
slow variation along $z$%
\begin{equation}
\Delta _{2}\Psi _{k}^{(i)}+\left( \frac{\omega _{0}^{2}}{c^{2}}\left( 1+\chi
^{(1)}\right) -\beta ^{(i)2}\right) \Psi _{k}^{(i)}=0  \label{B.7}
\end{equation}
$\left( i\right) $ denotes the mode number, $k$ the polarization and $\Delta
_{2}=\left( \frac{\partial ^{2}}{\partial x^{2}}+\frac{\partial ^{2}}{%
\partial y^{2}}\right) $.

Neglecting%
\footnote{justified for slow variations of the refractive index along
the beam axis over distances of the order of one wavelength}
$\frac{\partial ^{2}\Psi ^{(i)}}{\partial z^{2}}$ one obtains 
\begin{equation}
2i\beta ^{(i)}\frac{\partial \Psi _{1,2}^{(i)}}{\partial z}=-\frac{3\omega
_{0}^{2}}{4c^{2}}\chi ^{(3)}\left\{ \left( \left| \Psi _{1,2}^{(i)}\right|
^{2}+\frac{2}{3}\left| \Psi _{2,1}^{(i)}\right| ^{2}\right) \Psi
_{1,2}^{(i)}+\frac{1}{3}\Psi _{2,1}^{(i)}\Psi _{2,1}^{(i)}\Psi
_{1,2}^{(i)*}\right\}   \label{B.8}
\end{equation}
In directional couplers the propagating beams are made to overlap along one
of the transversal coordinates $\left( x\right) $. Typically, in the
nonlinear region of the directional coupler, the eigenmodes are symmetric $%
\left( +\right) $ and antisymmetric $\left( -\right) $ functions of $x$, the
amplitudes in each fiber at the input and output of the coupler being
recovered by 
\begin{equation}
\begin{array}{lll}
\Psi _{k}^{(1)} & = & \frac{1}{2}\left( \Psi _{k}^{(+)}+\Psi
_{k}^{(-)}\right)  \\ 
\Psi _{k}^{(2)} & = & \frac{1}{2}\left( \Psi _{k}^{(+)}-\Psi
_{k}^{(-)}\right) 
\end{array}
\label{B.9}
\end{equation}

An explicit analytic solution for the nonlinear coupler equation (\ref{B.8})
is now obtained by noticing that it has two constants of motion, namely 
\begin{equation}
\begin{array}{lll}
\frac{\partial }{\partial z}\left\{ \left| \Psi _{1}^{(i)}\right|
^{2}+\left| \Psi _{2}^{(i)}\right| ^{2}\right\}  & = & 0 \\ 
\frac{\partial }{\partial z}\left\{ \Psi _{1}^{(i)*}\Psi _{2}^{(i)}-\Psi
_{1}^{(i)}\Psi _{2}^{(i)*}\right\}  & = & 0
\end{array}
\label{B.10}
\end{equation}
Therefore, defining 
\begin{equation}
\begin{array}{lll}
\left| \Psi _{1}^{(i)}\right| ^{2}+\left| \Psi _{2}^{(i)}\right| ^{2} & = & 
\alpha ^{(i)} \\ 
\Psi _{1}^{(i)*}\Psi _{2}^{(i)}-\Psi _{1}^{(i)}\Psi _{2}^{(i)*} & = & 
i\gamma ^{(i)}
\end{array}
\label{B.11}
\end{equation}
one obtains for the electrical field of the eigenmodes 
\begin{equation}
\begin{array}{lll}
i\frac{\partial E_{1}^{(i)}}{\partial z} & = & -\stackrel{-}{\beta }%
^{(i)}E_{1}^{(i)}-i\stackrel{-}{k}^{(i)}E_{2}^{(i)} \\ 
i\frac{\partial E_{2}^{(i)}}{\partial z} & = & -\stackrel{-}{\beta }%
^{(i)}E_{2}^{(i)}+i\stackrel{-}{k}^{(i)}E_{1}^{(i)}
\end{array}
\label{B.12}
\end{equation}
with 
\begin{equation}
\begin{array}{lll}
\stackrel{-}{\beta }^{(i)} & = & \beta ^{(i)}+\frac{3\omega _{0}^{2}}{8c^{2}}%
\frac{\chi ^{(3)}}{\beta ^{(i)}}\alpha ^{(i)} \\ 
\stackrel{-}{k}^{(i)} & = & \frac{\omega _{0}^{2}}{8c^{2}}\frac{\chi ^{(3)}}{%
\beta ^{(i)}}\gamma ^{(i)}
\end{array}
\label{B.13}
\end{equation}
Notice that, through $\alpha ^{(i)}$ and $\gamma ^{(i)}$, $\stackrel{-}{%
\beta }^{(i)}$ and $\stackrel{-}{k}^{(i)}$depend on the material properties,
on the geometry of the mode and also on its intensity. One may now obtain,
for each eigenmode, the input-output relation of the nonlinear coupler 
\begin{equation}
\begin{array}{lll}
E_{1}^{(i)}\left( z\right)  & = & e^{i\stackrel{-}{\beta }^{(i)}z}\left\{
E_{1}^{(i)}\left( 0\right) \cos \left( \stackrel{-}{k}^{(i)}z\right)
-E_{2}^{(i)}\left( 0\right) \sin \left( \stackrel{-}{k}^{(i)}z\right)
\right\}  \\ 
E_{2}^{(i)}\left( z\right)  & = & e^{i\stackrel{-}{\beta }^{(i)}z}\left\{
E_{1}^{(i)}\left( 0\right) \sin \left( \stackrel{-}{k}^{(i)}z\right)
+E_{2}^{(i)}\left( 0\right) \cos \left( \stackrel{-}{k}^{(i)}z\right)
\right\} 
\end{array}
\label{B.14}
\end{equation}
($\left( i\right) =\left( +\right) $ or $\left( -\right) $), the
nonlinearity being embedded into $\stackrel{-}{\beta }^{(i)}$ and $\stackrel{%
-}{k}^{(i)}$%
\begin{equation}
\begin{array}{lll}
\stackrel{-}{\beta }^{(i)} & = & \beta ^{(i)}+\frac{3\omega _{0}^{2}}{8c^{2}}%
\frac{\chi ^{(3)}}{\beta ^{(i)}}\left( \left| E_{1}^{(i)}\left( 0\right)
\right| ^{2}+\left| E_{2}^{(i)}\left( 0\right) \right| ^{2}\right)  \\ 
\stackrel{-}{k}^{(i)} & = & \frac{\omega _{0}^{2}}{4c^{2}}\frac{\chi ^{(3)}}{%
\beta ^{(i)}}\textnormal{Im}\left( E_{1}^{(i)*}\left( 0\right) E_{2}^{(i)}
\left(0\right) \right) 
\end{array}
\label{B.15}
\end{equation}
To obtain the corresponding input-output relations in the two fibers $\left(
1\right) $ and $\left( 2\right) $ one uses Eqs.(\ref{B.9}), namely
\begin{equation}
\begin{array}{lll}
E_{k}^{(1)}\left( z\right)  & = & \frac{1}{2}\left( E_{k}^{(+)}\left(
z\right) +E_{k}^{(-)}\left( z\right) \right)  \\ 
E_{k}^{(2)}\left( z\right)  & = & \frac{1}{2}\left( E_{k}^{(+)}\left(
z\right) -E_{k}^{(-)}\left( z\right) \right) 
\end{array}
\label{B.16}
\end{equation}

In conclusion: Eqs.(\ref{B.14})-(\ref{B.16}) provide an analytic solution
for the nonlinear directional coupler, from which phases and polarizations
may be obtained explicitly.

\end{document}